\documentclass[preprint,authoryear,3p,times,12pt]{elsarticle}
\usepackage[utf8]{inputenc}
\usepackage{amsfonts,amsmath,bm,bbm}
\usepackage{amsmath}
\usepackage{graphicx}
\usepackage{epstopdf}
\usepackage{amssymb}
\usepackage{mathtools}
\usepackage{enumerate}
\usepackage{natbib}
\usepackage{indentfirst}
\usepackage[font=footnotesize]{caption}
\usepackage{floatrow}
\usepackage{enumitem}
\usepackage{dsfont}
\DeclareMathOperator{\argmin}{argmin}
\usepackage{subcaption}

\usepackage{booktabs}
\usepackage{hyperref}

\usepackage{color}
\definecolor{gray}{rgb}{0.5,0.5,0.5}
\definecolor{dgreen}{rgb}{0,0.5,0}
\long\def\red#1{{\color{black}{#1}\color{black}}}

\journal{International Journal of Forecasting}

\begin{document}
	\begin{frontmatter}
		
		\title{LASSO Principal Component Averaging -- a fully automated approach for point forecast pooling}
		
		\author[KBO]{Bartosz Uniejewski}
		\ead{bartosz.uniejewski@pwr.edu.pl}
		\author[KBO]{Katarzyna Maciejowska}
		\ead{katarzyna.maciejowska@pwr.edu.pl}
		
		\address[KBO]{Department of Operations Research and Business Intelligence, Wroc{\l}aw University of Science and Technology, 50-370 Wroc{\l}aw, Poland}
		
		\date{This version: \today}

		\begin{abstract}
		\red{This paper develops a novel, fully automated forecast averaging scheme, which combines LASSO estimation method with Principal Component Averaging (PCA). LASSO-PCA (LPCA) explores a pool of predictions based on a single model but calibrated to windows of different sizes. It uses information criteria to select tuning parameters and hence reduces the impact of  researchers' \emph{at hock} decisions.  The method is}
		applied to average predictions of hourly day-ahead electricity prices over 650 point forecasts obtained with various lengths of calibration windows. It is evaluated on four European and American markets with almost two and a half year of out-of-sample period and compared to other semi- and fully automated methods, such as simple mean, AW/WAW, LASSO and PCA. The results indicate that the LASSO averaging is very efficient in terms of forecast error reduction, whereas PCA method is robust to the selection of the specification parameter. LPCA inherits the advantages of both methods and outperforms other approaches in terms of MAE, remaining insensitive the the choice of a tuning parameter.
		
		
		
		\end{abstract}

		
	\end{frontmatter}
	
	\section{Introduction}

\emph{Electricity price forecasting} (EPF) is nowadays \red{perceived as} fundamental \red{for} decision making in energy \red{markets}. As short-term transactions provide a tool for adjusting long-term positions and \red{a} benchmark in over-the-counter (OTC) trading, the day-ahead, intraday and balancing prices play a key role in day-to-day operations \citep{kat:zie:18,mac:nit:wer:19,may:tru:18,wer:14}. In the last decades, the market share of renewable energy sources has rapidly increased. \red{As a result, intermittent} changes in generation \red{level and structure} have become more likely to occur. \red{This leads to an increase  of market imbalances and a rise of electricity price volatility} \citep{gia:par:pel:16,k-p:18, mac:20}. \red{Hence},  reliable methods \red{dedicated} to \red{EPF} are more than essential in \red{rational managing of} energy companies. 

One of the methods to increase the prediction accuracy is to combine forecasts obtained with different models. The idea of \red{forecast} averaging has started about half a \red{century} ago. Pioneering papers of \cite{bat:gra:69} and \cite{cra:cro:67} inspired many authors to develop \red{new methods and contribute to} the area. Since the late 60s, hundreds of papers have suggested the superiority of forecast combinations over individual models \citep{tim:06, wal:11, now:wer:16:ARGO}. \cite{hib:evg:05} 
\red{states} that the main advantage of combining forecasts is the fact that, in practice, it is less risky to combine forecasts than to select an individual forecasting method. 

Recently, more and more experts \red{have} put \red{an} attention to a selection of calibration window, \red{which is} used \red{for model} estimation \red{\cite[see][]{pes:tim:07}}. \cite{mar:bun:bel:ren:20} claim \red{that} in rapidly developing markets, \red{such as an} energy market, \red{researchers} should take into account structural breaks and adjust \red{model} parameters to \red{market} changes. The simplest solution \red{of the issue} is to work \red{with} short data, \red{which describes only the most recent events}. This approach \red{has some severe drawbacks} as it \red{decreases the estimation accuracy and limits the complexity of applied models}. On the other hand, \red{one may} try to \red{estimate a time of a} structural break  and include it \red{directly} in a forecasting model. \red{An assumption of  a discrete shift of model parameters is however not suitable for more complex evolution patterns} \citep{mar:bun:bel:ren:20}. \red{In the literature, there is no agreement, which solution is the best and therefore} the majority of research in \red{EPF applies} an arbitrary chosen calibration window length. 
In \red{recent articles} , \cite{mar:ser:wer:18,hub:mar:wer:19,ser:uni:wer:19} \red{suggest} to use a pool of different in-sample data sizes and to average \red{the resulting} forecasts. The \red{outcomes presented in these papers}  suggest that \red{a choice of} three 'short' and three 'long' calibration windows \red{provide robust results, which outperform all individual predictions}. \red{This conclusion is questioned by } \cite{mac:uni:ser:20}, \red{who show that} the suggested \red{solution is not valid for} all the electricity markets and has to be adjusted to a market specification. 

The estimation of a single model with various calibration windows enables to obtain a large number of predictions. \red{For example, in \cite{mac:uni:ser:20} a panel of  673 forecasts is built. Moreover, it could be observed that predictions in such a pool are very similar to each other, because a slight change of an estimation window does not alter much the model parameters}. Thus, it is natural to search for methods that \red{would} help to reduce the dimension of the problem, without losing useful information. In this context, two approaches are natural candidates: Principal Component (PC) method, which summarizes the panel with a small number of components  \citep[see][]{sto:wat:02a, bai:ng:02} and Least Absolute Shrinkage and Selection Operator \red{\cite[LASSO, ][]{tib:96}}, which reduces the dimension of a model by assigning a penalty to non-zero parameters. Here, \red{we propose a novel approach, which combines these two methods and } apply them to forecast averaging.

\red{PC} is a well-known tool, \red{which has been} successfully applied for analyzing big panels of data. It has been used to predict directly the variables of interest \citep{boi:ng:05, sto:wat:12} or to augment a small-scale econometric model \citep{ban:mar:mas:14}. The factor models has been extended to account for dynamic relationships \citep[see][]{for:hal:lip:rei:00, for:lip:01} and used to create economic indicators \citep{sto:wat:98}. Although  
the potential of PCA in forecast averaging area was recognized by \cite{cha:sto:wat:1999} and \cite{hua:lee:10}, there are only few papers, which illustrate its performance. \cite{sto:wat:04} and \cite{pon:rod:san:sen:11} used PCA to predict macroeconomic variables. They estimated components from a panel of forecasts coming from  either different models or different experts. In both cases, the panels were relatively small and diversified. \cite{mac:uni:ser:20} proposed an algorithm, which extracts PCs from a standardized, large panel of predictions coming from a single model (as in \cite{mar:ser:wer:18}, \cite{hub:mar:wer:19} and \cite{ser:uni:wer:19}) and use them to calculate the final forecasts via linear regression. In this article, 1--4 components were used. The results indicate that PCA is a robust method for forecast pooling. The major issue of \cite{mac:uni:ser:20} is the fact that the number of PCs is either chosen a prior or selected from a small number of alternatives. Moreover, it is not clear how the approach will perform, if a larger number of components is considered.

\red{The literature proposes many methods of dealing with a large set of potential explanatory variables. Two major approaches could be distinguished: selecting an optimal model  \citep{lud:feu:neu:15,zie:ste:hus:15,gai:gou:ned:16,uni:now:wer:16} or average across models (\cite[][]{yan:01, han:rac:12,wan:pat:gao:yan:14}). Here, we adopt the first approach and apply LASSO}, which was introduced by \cite{tib:96} and \red{is one of the most popular and important regularization methods.} Because of \red{its} linear penalty function, the LASSO estimator shrinks the coefficients of the less important explanatory variables \red{to} zero. It becomes a tool for automated variable selection, as it identifies the \red{significant} variables and excludes the redundant ones \citep{uni:now:wer:16, uni:wer:18}. In the context of prediction pooling, the LASSO technique has been successfully used in both point \cite{die:shi:19} and probabilistic \cite{bay:18, bra:car:def:19,uni:wer:21} forecasting. It is worth noticing that, to our best knowledge, LASSO averaging has not been applied to point forecasting of electricity prices and therefore there is a need to evaluate its performance in this field.

The main novelty of this paper is a fully automated forecast averaging scheme, which utilizes both PCA and LASSO regularization techniques. We present an algorithm, which extends the approach described in \cite{mac:uni:ser:20} and allows to use an arbitrary large number of components. Thanks to LASSO estimation method, the irrelevant PCs are excluded and hence the corresponding noise is reduced. Since LASSO depends on a tuning parameter, Information Criteria (IC) are applied to select its optimal value. Unlike in a typical LASSO averaging, the inputs in LPCA are orthogonal to each other. Moreover, although one could use all PCs, a smaller number of components than individual forecasts should be sufficient. Hence, LPCA should be much easier and faster to compute than the full panel LASSO. 
As a result, the proposed methodology does not require any expert knowledge nor intuition to obtain the prediction of future prices but also should be less computationally burdensome  than existing methods.

The paper is structured as follows. First, we present the datasets that consist of day-ahead price series as well as exogenous variables. At the end of section \ref{sec:data} we describe a data transformation. Next, in Section \ref{sec:methodology} we present the methodology, first, to obtain point forecasts and afterwards to average them. In the same section, we introduce a new algorithm for a fully automated approach designed to combine forecasts. Finally, in Section \ref{sec:res} we present the results of our study and in Section \ref{sec:conclusion} we conclude the research. 

\section{Datasets}
\label{sec:data}

The datasets used in this study cover five years and \red{describe} four different markets: German (EPEX), Scandinavian (Nord Pool, NP), Spanish (OMIE) and American (PJM).
All time series have an hourly resolution and span 1826 days
from 1.01.2015 to 31.12.2019 (the data is not extended to 2020, as the COVID-19 pandemic has changed the market dynamics). The missing or 'doubled' values (corresponding to the time change) \red{are} replaced by the average of the closest observations, for the missing hours, and the arithmetic mean of the two values, for 'doubled' hours. Note that the data is double indexed, with $d$ denoting the day and $h$ the hour of an observation.

\subsection{Day-ahead electricity prices}
This research focuses on electricity prices from day-ahead markets, which are established simultaneously around noon on the day preceding the delivery. A more detailed description of the day-ahead market design can be found in \cite{wer:14}. As a result, market participants can utilize only the information available at the time of bidding. This impacts also the forecasters, who should include in their models only the data published before the noon \citep[see][]{hui:huu:mah:07}.

\red{In this article, the following day-ahead prices, $DA_{d,h}$, are considered}:
\begin{itemize}
    \item the German market EPEX spot (\emph{top} panel in Figure \ref{fig:EPEX}); the data taken from transparency platform (\url{https://transparency.entsoe.eu})
    \item the Scandinavian market Nord Pool (\emph{top} panel in Figure \ref{fig:NP}); the data taken from Nord Pool website (\url{https://www.nordpoolgroup.com)})
    \item the Spanish market OMIE (\emph{top} panel in Figure \ref{fig:OMIE}); the data taken from OMIE website (\url{https://www.omie.es})
    \item the American market PJM COMED (\emph{top} panel in Figure \ref{fig:PJM}); the data  taken from PJM data miner (\url{https://dataminer2.pjm.com})
\end{itemize}

\subsection{Exogenous variables}
\label{ssec:exogenous}
The literature indicates that various exogenous factors, such as generation structure or fuel prices, have an important impact on the electricity prices and can be used for their forecasting \red{\cite[][]{gia:rav:ros:20,bil:gia:gro:rav:22}}. Following \cite{mac:uni:ser:20}, \red{in this study,} we consider day-ahead predictions of fundamental variables describing the demand and supply of electricity, \red{which are} provided by transmission system operators (TSO). The description of the data can be found in Table \ref{tab:exogenous variables}. \red{Notice that the set of exogenous variables changes between markets and depends on the data availability.}

\begin{table}[H]
	\centering
 	\caption{Exogenous variables}
	\label{tab:exogenous variables}
	\begin{tabular}{lcc}
		\toprule
			Description & Notation & Availability \\
		\midrule
 Load & $L_{d,h}$ & EPEX, NP, OMIE, PJM\\
Zonal load &$Z_{d,h}$ &  PJM\\
 Wind power generation (WPG) &$W_{d,h}$ & EPEX, NP, OMIE\\
Photovoltaic generation (PVG) &$S_{d,h}$ &  EPEX, OMIE\\
\toprule
		\end{tabular}
\end{table}

\begin{figure}[p]
     \centering
     
     \begin{subfigure}[b]{0.75\textwidth}
         \centering
         \includegraphics[width=\textwidth]{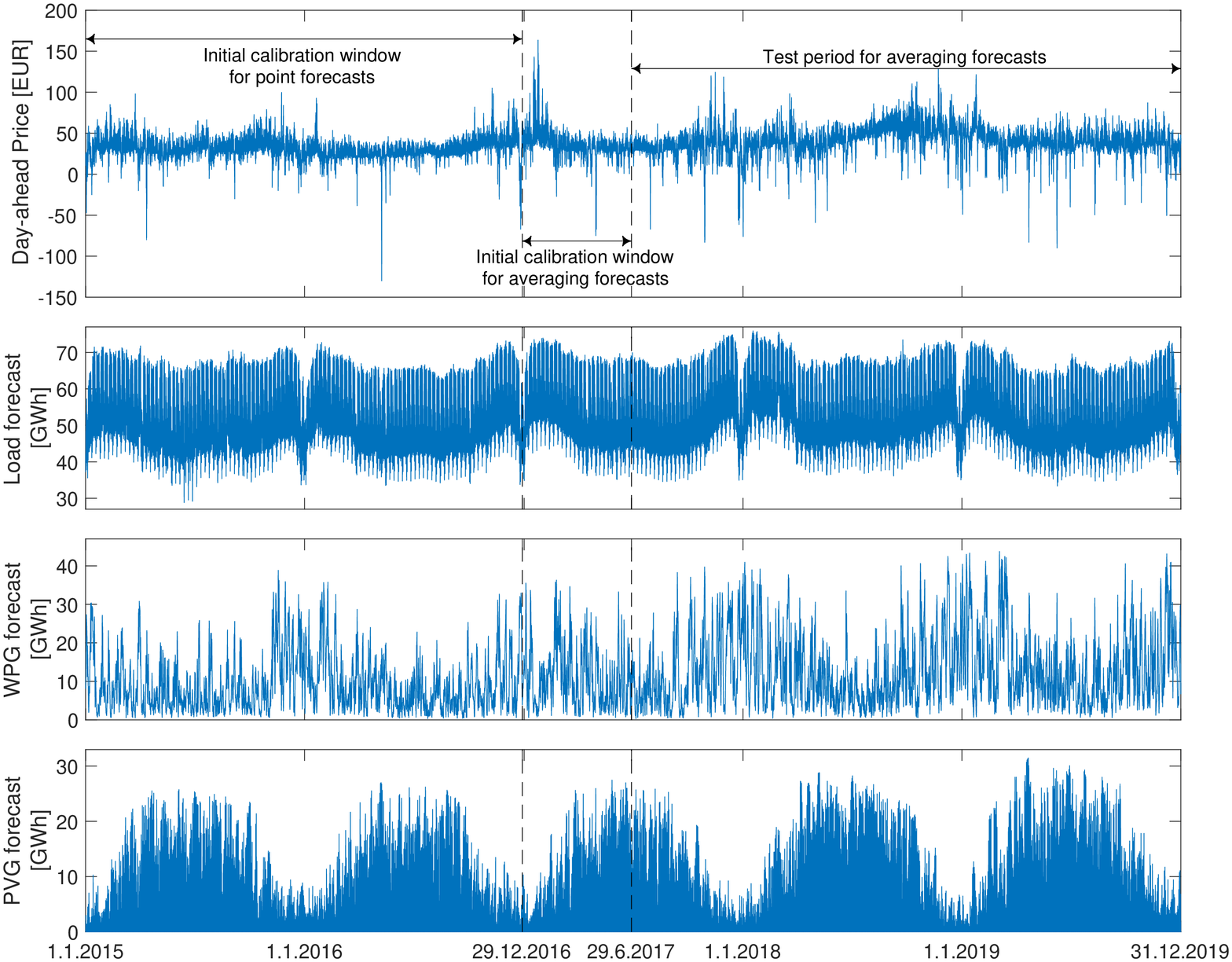}
         \caption{EPEX system prices (\textit{top}), day-ahead consumption prognosis (\textit{middle top}), day-ahead forecasts of wind power generation (\textit{middle bottom}),day-ahead forecasts of Photovoltaic generation (\textit{bottom})}
         \label{fig:EPEX}
     \end{subfigure}
     
     \vspace{0.3cm}
     
     \begin{subfigure}[b]{0.75\textwidth}
         \centering
         \includegraphics[width=\textwidth]{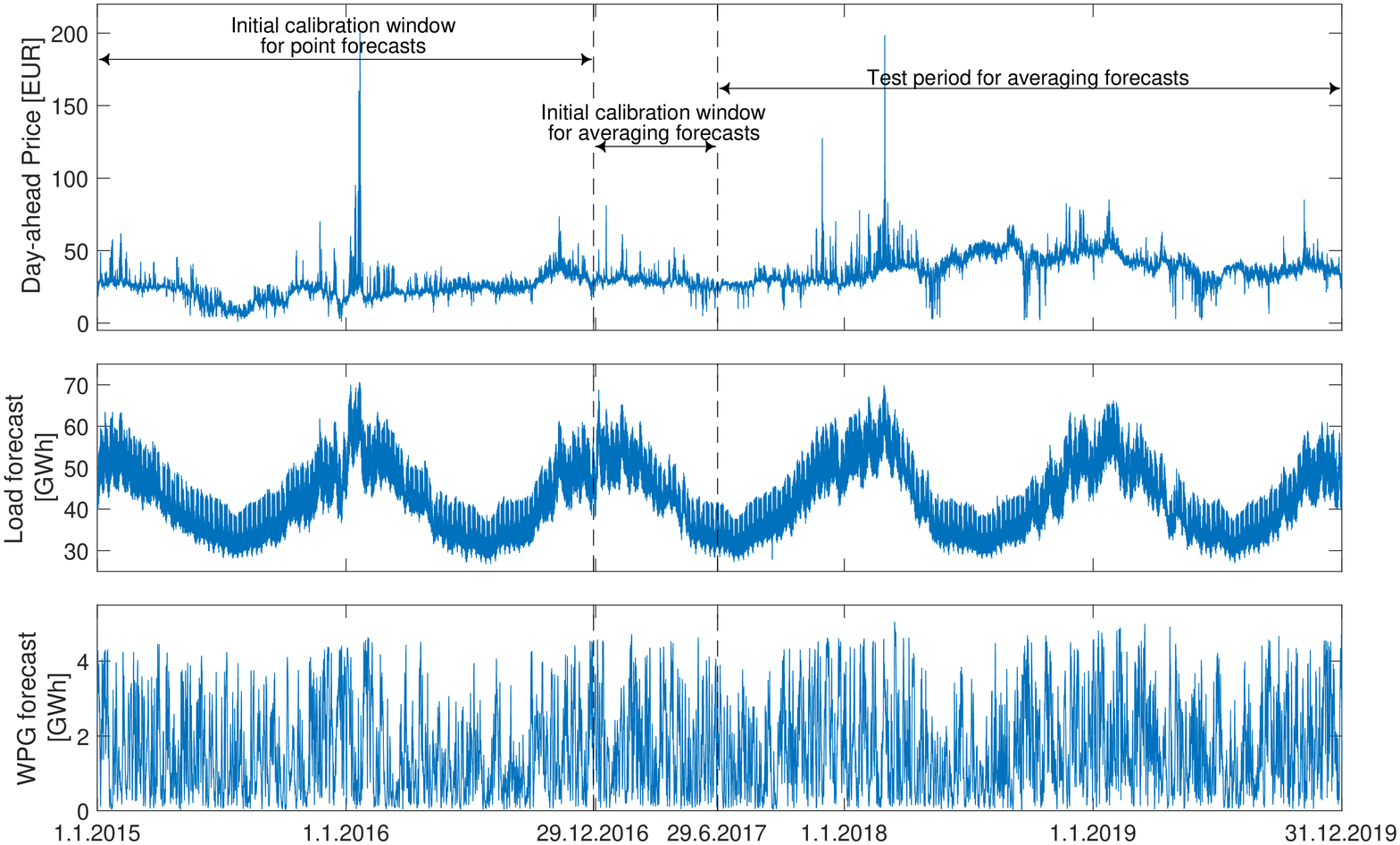}
         \caption{Nord Pool system prices (\textit{top}), day-ahead consumption prognosis (\textit{middle}), day-ahead forecasts of wind power generation (\textit{bottom})}
         \label{fig:NP}
     \end{subfigure}
     \caption{Day-ahead prices and exogenous time series from 1 January 2015 to 31 December 2019. The vertical dashed lines mark respectively the beginning of the out-of-sample test period for point forecasts (29 December 2016; also the beginning of the initial 182-day calibration window for averaging forecasts) and the beginning of the out-of-sample test period for averaging forecasts (29 June 2017). The first 728 days constitute the initial calibration window for point forecasts.}
     \label{fig:data1}
\end{figure}

\begin{figure}[p]
     \centering
     
     \begin{subfigure}[b]{0.75\textwidth}
         \centering
         \includegraphics[width=\textwidth]{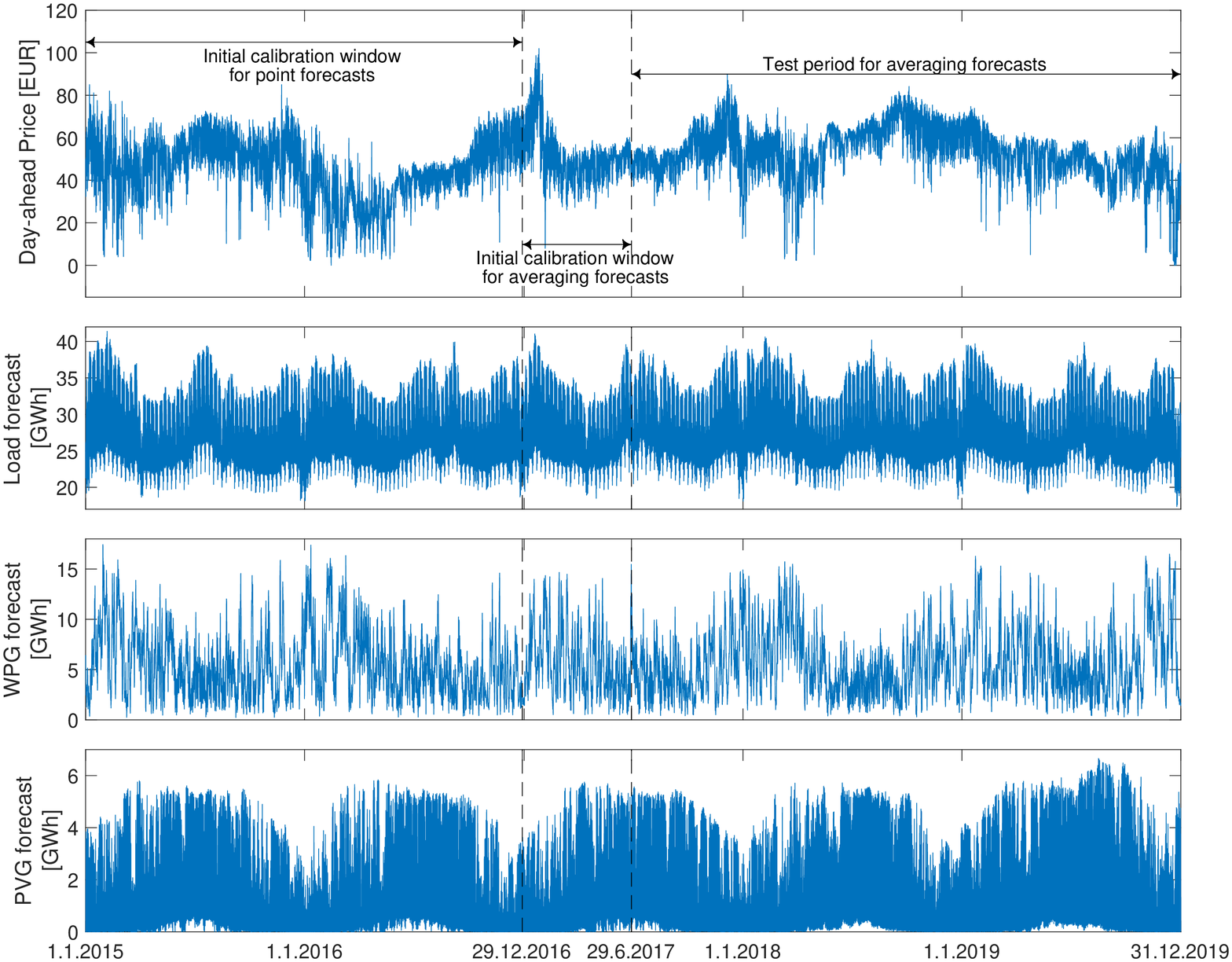}
         \caption{OMIE system prices (\textit{top}), day-ahead consumption prognosis (\textit{middle top}), day-ahead forecasts of wind power generation (\textit{middle bottom}),day-ahead forecasts of Photovoltaic generation (\textit{bottom})}
         \label{fig:OMIE}
     \end{subfigure}
     
     \vspace{0.3cm}
     
     \begin{subfigure}[b]{0.75\textwidth}
         \centering
         \includegraphics[width=\textwidth]{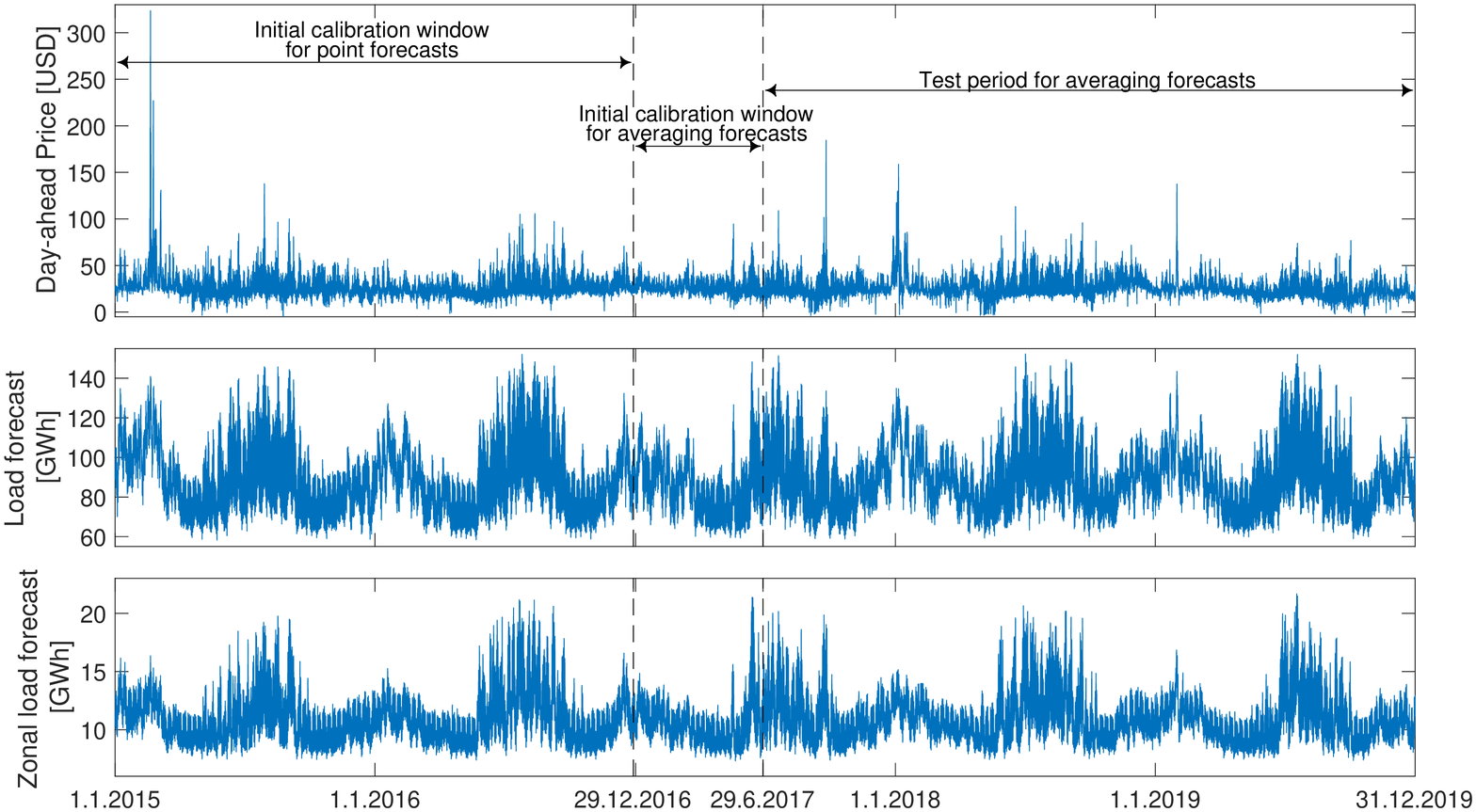}
         \caption{PJM system prices (\textit{top}), day-ahead system load prognosis (\textit{middle}), day-ahead zonal (COMED) load prognosis (\textit{bottom})}
         \label{fig:PJM}
     \end{subfigure}
     \caption{Day-ahead prices and exogenous time series from 1 January 2015 to 31 December 2019. The vertical dashed lines mark respectively the beginning of the out-of-sample test period for point forecasts (29 December 2016; also the beginning of the initial 182-day calibration window for averaging forecasts) and the beginning of the out-of-sample test period for averaging forecasts (29 June 2017). The first 728 days constitute the initial calibration window for point forecasts.}
     \label{fig:data2}
\end{figure}

The day-ahead forecasts for all exogenous variables are plotted in Figures \ref{fig:data1} and \ref{fig:data2}.  The  variables, in particular load and solar generation, exhibit strong yearly seasonality, with the load following also a weekly pattern.

\subsection{Variance Stabilizing Transformation}
As it can be easily seen in Figures \ref{fig:data1} and \ref{fig:data2}, electricity prices exhibit spiky behavior.
\citet{uni:wer:zie:18} argue that it is possible to reduce the influence of such extreme values on forecasts by using a Variance Stabilizing Transformation (VST). These findings are confirmed by the literature \citep{uni:wer:18,mar:ser:wer:18}. Here, we follow the recommendation of \cite{uni:wer:zie:18} and apply the N-PIT transformation (to all variables in the dataset).
Let us recall that the N-PIT transformation is based on the so-called \emph{probability integral transform}. Lets consider a time series $Y_{d,h}$. Its transformation, \red{$\tilde{Y}_{d,h}$}, is given by:
\begin{equation}
\red{\tilde{Y}}_{d,h} = N^{-1}\left( \hat{F}_{Y}(Y_{d,h}) \right),
\end{equation}
where $\hat{F}_{Y}(\cdot)$ is the empirical cumulative distribution function of in-sample $Y$, and $N^{-1}$ is the quantile function of normal distribution.
After the models are estimated on the transformed time series, we apply the inverse transformation to obtain the final forecast of electricity price:
\begin{equation}
\red{Y}_{d,h} = \hat{F}_{Y}\left( N(\red{\tilde{Y}}_{d,h}) \right),
\end{equation}
where the time series $Y$ corresponds to price series $DA$.

\section{Methodology}
\label{sec:methodology}

\subsection{Experiment design}
\label{ssec:CalibrationWindows}

The majority of research in \red{EPF} literature \red{chooses} arbitrary the \red{length of a} calibration window. \red{In last years, various research \cite[see][]{mar:ser:wer:18,hub:mar:wer:19,ser:uni:wer:19,mac:uni:ser:20} has shown that averaging predictions based on different in-sample data leads to an improvement of the forecast accuracy. Here,}
 we follow \red{this} idea and use a pool of 673 calibration window lengths -- ranging from 56 (ca.\ two months) to 728 days (ca.\ two years). Unlike in previous papers, this research focuses on the automatisation of the averaging process  \red{in order} to make it independent of \red{ \emph{at hock} decisions} of forecasters.
 
\red{The pool of forecasts is obtained with a rolling window procedure, a standard procedure in EPF literature \citep[][]{wer:14}.}
 To be more specific, the first 728 days \red{are used} for model estimation (for shorter windows, the calibration sample is left truncated, so it ends on the same day). Next, 24 point forecasts \red{are computed}, one for each hour of the day, and finally the window \red{is moved} one day forward. The procedure is repeated until the last out-of-sample day \red{is reached}. 
Once the pool of predictions is created, a rolling  window of 182 days (ca.\ half of a year) \red{is used} to calibrate the averaging methods (see Section \ref{ssec:averaging}). 
The final predictions are evaluated using the last 916 days of the sample. The division into the point forecast, averaging and the out-of-sample \red{periods} are marked by dashed lines in Figures \ref{fig:data1} and \ref{fig:data2}. The first line marks the end of the initial 728-day calibration window for point forecasts (i.e., 1 January 2015 to 28 December 2016). The second indicates the end of the initial 182-day calibration window for averaging forecasts (i.e., 28 June 2017), which is also the beginning of the evaluation period.

\subsection{Forecasting models}

In this research, \red{forecasts} for all 24 hours of the next day \red{are} computed simultaneously a day in advance. Similarly to \cite{mac:uni:ser:20}, we consider a parsimonious autoregressive structure used in a number of EPF studies \citep{uni:now:wer:16, zie:wer:18, uni:wer:zie:18, uni:wer:18}. The originally proposed setup is expanded to include the exogenous variables presented in Section \ref{ssec:exogenous}. The final model is denoted by $\mathbf{DA}$. The price $DA_{d,h}$ for day $d$ and hour $h$ is described by the following formula:

\begin{align}
\label{eqn:DA_model}
DA_{d,h} = &~\underbrace{ \beta_{h,1}  DA_{d-1,h} + \beta_{h,2} DA_{d-2,h} + \beta_{h,3} DA_{d-7,h} }_{\scriptsize \mbox{autoregressive effects}} + \nonumber \\
& + \underbrace{ \beta_{h,4} DA_{d-1,min} + \beta_{h,5} DA_{d-1,max}}_{\scriptsize \mbox{non-linear effects}} + \underbrace{ \beta_{h,6}  DA_{d-1,24} }_{\scriptsize \mbox{midnight price}} + \nonumber \\
& + \underbrace{ \sum_{i=1}^{7}\beta_{h,6+i} D_i }_{\scriptsize \mbox{weekday dummies}} + \underbrace{ \theta_{h} X_{d,h} }_{\scriptsize \mbox{exogenous variables}}+  \varepsilon_{d,h},
\end{align}
where $DA_{d-1,h}$, $DA_{d-2,h}$, $DA_{d-7,h}$ are the lagged day-ahead prices from one, two and seven days before. $DA_{d-1,min}$ and $DA_{d-1,max}$ refer  to the minimum and the maximum price from day $d-1$, \red{respectively.} $DA_{d-1,24}$ is the last already known price, corresponding to the previous day midnight.  \red{$D_1,\ldots,D_7$ denotes dummies, which capture weekly seasonality}. 
\red{Finally, the vector $X_{d,h}$ describes exogenous variables. As stated in Section \ref{ssec:exogenous}, $X_{d,h}$ differs across markets.} The day-ahead forecasts of the load ($L_{d,h}$) \red{are} included \red{in $X_{d,h}$} for all the countries\red{, whereas presence of other variables is restricted by their availability. For example, $W_{d,h}$ is used for all European country but is not included in $X_{d,h}$ for PJM market}. Therefore, \red{in  case of USA,}  the zonal load forecasts ($Z_{d,h}$) \red{are added instead}. Additionally, for Germany and Spain, the photovoltaic generation $S_{d,h}$ \red{is included}. Note that, as in \cite{mac:nit:wer:19} and \cite{mac:uni:ser:20}, \red{$S_{d,h}$ is admitted in the model(\ref{eqn:DA_model})} only for hours 9-17 because  during the night and early morning hours \red{the solar generation is too weak to impact the electricity price}.

\subsection{Averaging methods}
\label{ssec:averaging}

According to recent literature, the forecasting performance of statistical models is sensitive to the choice of the calibration window \red{\cite[][]{hub:mar:wer:19}}. \red{Hence}, it may be beneficial to average forecasts based on windows of different lengths \citep{pes:tim:07,hub:mar:wer:19}
\red{as it allows to explore both a local and a long-run behavior. Although estimation of the same model with different data sets seems straightforward, the forecast averaging remains a demanding task}.
First, it could be noticed that a large number of predictions, which are based on long windows, are almost identical. Extending the sample by one observation from, for example, 727 to 728 days, does not alter much the parameter estimates. This feature impedes the usage of typical regressions for choosing averaging weights, as a large number of forecasts are almost co-linear. On the other hand, there is a relatively small number of predictions based on short windows, which are distinct. Unfortunately, these forecasts are also more variable and typically burdened with a larger forecast error. Finally, it is not clear how to balance the impact of the short- and long- windows on the final prediction.

In this paper, we consider three types of forecast combining methods. First, predictions are computed  either as a simple or as a weighted  mean of individual forecasts. Next, the weights are selected with LASSO method, which is a regression-based approach. LASSO allows to include a large number of input variables and shrinks the parameters toward zero. Hence, it can help to select the optimal window lengths. Finally, the information included in the panel of forecasts is summarized by a set of common factors (computed as principal components, PCs), which are next used to compute the predictions of interest. 

\subsubsection{Linear average (simple average, AW, WAW)}
\label{ssec:AW}

In this research, we consider three methods based on a linear average. The literature indicates that the arithmetic mean is a simple but very efficient approach  \citep{gen:ken:mey:tim:13}. Here, we compute the mean of all considered windows' sizes ranging from 56 days to 728 days and denote it by a \textbf{simple average}. Second, following \cite{hub:mar:wer:19}, a subset of six calibration window lengths is selected, which consists of three short (56-, 84-, 112-days) and three long (714-, 721, 728-days) in-sample sizes. Forecasts based on these chosen window sizes are next averaged. This approach is denoted  by \textbf{AW(56, 84, 112, 714, 721, 728)}) or simply \textbf{AW}. Unfortunately, both: simple average and AW, assume that the weights are equal and constant over time. Therefore, they cannot adopt to changing market conditions, for example, a rising share of renewable energy sources (RES) in the generation mix.

In order to overcome this problem, \cite{mar:ser:wer:18} proposed to extend AW to allow for data-driven weights. Similar to \cite{hub:mar:wer:19}, a small subset of available forecasts is  first selected. Then, instead of taking a simple average, \cite{mar:ser:wer:18} use the forecast errors from the previous day to assign weights to each individual prediction. The forecasts are evaluated with Mean Absolute Error (MAE) and those, which are more accurate, get higher weights (for more details see equation (5) in \cite{mar:ser:wer:18}). Here, following \cite{mac:uni:ser:20}, we use the whole averaging window (182 days) to compute the weights. Similar to AW, the Weighted AW is denoted as  \textbf{WAW(56, 84, 112, 714, 721, 728)} or simply \textbf{WAW}.

An application of linear averages is associated with some issues. First, when computing the simple average, the majority of inputs come from long calibration windows, which provide very similar forecasts. Hence, the long windows dominate and reduce the impact of local behavior. This drawback is reduced in AW and WAW approaches, as they include the same number of short and long windows and balance the impact of different window sizes. Unfortunately, AW/WAW, unlike the simple average, requires pre-selection of the number and lengths of calibration windows used for averaging. Hence, it could not be considered as a robust approach because a subset, which works well for one market, may not be plausible for the other.

\subsubsection{LASSO averaging}

The idea of regularization \red{of an estimation process} can be \red{viewed} as an optimization problem:
\begin{equation}
\label{regularization}
\hat{\boldsymbol\beta} = \argmin \left\{f(\boldsymbol X;\boldsymbol\beta) + g(\boldsymbol\beta)\right\},
\end{equation}
where \red{$\boldsymbol\beta $ is a parameter vector and $\boldsymbol X$ is a data set. In equation (\ref{regularization}), $f(\boldsymbol X;\boldsymbol\beta)$ denotes a loss } function, e.g. the Residual Sum of Squares \red{(RSS) as in Least Squares estimation method}, while $g(\boldsymbol \beta)$ is the penalty function \cite[][]{tik:63}.

\red{In the literature, it is common to use a scaled $\ell^q$ norm as $g(\boldsymbol \beta)$. The most popular variant of the regularization, called LASSO, was introduced by \cite{tib:96}. It sets $q=1$ and $f(\boldsymbol X;\boldsymbol\beta) = \text{RSS}$ (see \eqref{eqn:LASSO}).}
 \red{Due to its properties, it} becomes a tool for automated variable selection and can successfully identify the most important variables \citep{uni:now:wer:16, uni:wer:18}. 

\begin{equation}\label{eqn:LASSO}
\hat{\boldsymbol\beta} 
= \argmin \left\{ \mbox{RSS} + \lambda \sum_{i=1}^{n} |\beta_{h,i}| \right\}
\equiv \argmin \left\{ \sum_{d,h} \left( p_{d,h} - \sum_{i=1}^{n} \beta_{h,i} X_{d,h,i} \right)^2 + \lambda \sum_{i=1}^{n} |\beta_{h,i}| \right\},
\end{equation}

LASSO is \red{also} one of the most popular solutions to combine point forecasts. It becomes a gold standard in the literature, especially for high-dimensional problems (it is when the number of individual predictions exceeds the number of in-sample observations). It has the property of selecting only a few individual point forecasts even in the case of rich pools, which benefits in accuracy improvement. In a recent paper, \cite{uni:wer:21} showed that the linear penalty regularization works also in probabilistic forecasting. 

In this article, LASSO regression is used to average all (673) point forecasts from the pool (see Section \ref{ssec:CalibrationWindows}). We consider a log-scaled grid of 20 $\lambda$ parameters (\textbf{LASSO($\lambda$)}) and choose its optimal value via Information Criteria: AIC, BIC and HQC. The procedure to select the tuning parameter is taken from \cite{zie:wer:18} and its results are denoted by \textbf{LASSO(BIC)}, \textbf{LASSO(AIC)}, \textbf{LASSO(HQC)}.
\label{ssec:LASSO}

\subsubsection{Principal Component Averaging (PCA)}
\label{ssec:PCA}
\red{Many} forecast averaging methods strongly depend on expert knowledge. For example, AW and WAW require pre-selection of window lengths used in the forecast pooling. In order to overcome this issue, \cite{mac:uni:ser:20}  proposed to use Principal Component Averaging (PCA) to automate the procedure of averaging over a rich pool of predictions. Authors applied the principal component method to a panel of over 650 point forecasts obtained with models calibrated with different in-sample sizes. Next, they used the estimated components in a linear regression to form the final predictions. In such a way, they overcome the problem of co-linearity of forecasts stemming from the same model calibrated on similar windows.  Their results indicated that the PCA forecast averaging leads to more accurate predictions of electricity prices in terms of MAE than the simple average, AW or WAW.

\red{The step-by-step algorithm of PCA is described below. In the algorithm, $d_f$ denotes the forecasted day and $\tau = 56, 57, \ldots, 728$ stands for the length (in days) of a calibration window used to calculate the predictions. Moreover, during the averaging, all the hourly predictions are treated as time series and indexed with $t$. The averaging window includes the predicted day $d_f$ and 182 proceeding days:  $t \in \{24d+h: d_f-182\leq d \leq d_f, 1\leq h \leq 24\}$. 
Finally, in the following parts of the paper, $\hat{P}_{t,\tau}$ denotes the predicted electricity prices for period $t$ obtained with a $\tau$-day calibration window, whereas $P_t$ stands for their actual level. }

\begin{enumerate}
	\item For each time \red{period, $t$,  in averaging window} estimate the mean ($\hat{\mu}_{t}$) and standard deviation ($\hat{\sigma}_{t}$) of forecasts ($\hat{P}_{t,\tau}$) across $\tau = 56, 57, \ldots, 728$.
	\item Standardize forecasts and the real price with \red{previously} estimated  $\hat{\mu}_{t}$ and $\hat{\sigma}_{t}$:
	\begin{equation}
	\hat{Z}_{t,\tau}=\frac{\hat{P}_{t,\tau}-\hat{\mu}_{t}}{\hat{\sigma}_{t}}, \quad 
	Z_{t}=\frac{P_{t}-\hat{\mu}_{t}}{\hat{\sigma}_{t}}.
	\end{equation}
	\red{Notice that at the time of forecasting, the last 24 elements of $Z_t$, corresponding to the predicted day $d_f$, are not known.}
	\item Estimate the first $K$ principal components, \red{($PC_{t,1}, PC_{t,2}, \ldots. PC_{t,K}$)},  of a panel $\{\hat{Z}_{t,\tau}\}$, using the method described by \cite{bai:ng:02,sto:wat:04}.  \red{Notice that PCs } include the information of the price forecasts for all hours in 182-days long averaging calibration window as well as the forecasted day.
	\item Estimate linear regression parameters with Least Squares (LS)  using observations from the averaging window (without day $d_f$)
	\begin{equation}
	\label{eq:reg:P:PC}
	{Z}_{t}=\alpha+\sum_{k=1}^K\beta_kPC_{t,k}+\varepsilon_{t}
	\end{equation}

	\item Using estimated parameters compute the prediction of the normalized price $Z_{t}$ for $t \in \left( 24d_f+1, 24d_f+24\right) $ corresponding to all hours in forecasted day $d_f$:
	\begin{equation}
	\label{eq:reg:P:forecast}
	\hat{Z}_{t}=\hat{\alpha}+\sum_{k=1}^K\hat{\beta}_kPC_{t,k}
	\end{equation}
	and transform it back into its original level
	\begin{equation}
	\label{eq:reg:P:transform}
	\hat{P}_{t}=\hat{Z}_{t}\cdot\hat{\sigma}_{t}+\hat{\mu}_{t}
	\end{equation}
	
\end{enumerate} 

Although PCA allows to explore the information included in the whole panel of forecasts, it still requires selection of the number of components used in a regression, $K$. Therefore, similar to \cite{mac:uni:ser:20}, we consider the method based on $k$-first PCs and denote them by \textbf{PCA($k$)}. For illustrative purposes, we  also choose ex-post optimal (fixed) number of PCs taken for averaging and denote it by \textbf{PCA(best)}.

Next, three variants of PCA are applied, which are based on Information Criteria (IC). This allows a data-driven adjustment of the number of PCs used in the regression (\ref{eq:reg:P:PC}). We consider the same ICs, which are used to select $\lambda$ in LASSO procedure. The results are denoted  consecutively by \textbf{PCA(BIC)}, \textbf{PCA(AIC)}, \textbf{PCA(HQC)}.

\subsubsection{LASSO Principal Component Averaging (LPCA)}

In this paper, we propose a novel approach, which combines PCA-based procedure with LASSO estimation method. First, similar to  \cite{mac:uni:ser:20}, $K$ components are extracted from the standardized  panel of point predictions (see Section \ref{ssec:PCA} for a detailed description of the algorithm). Unlike in previous work, the number of PCs is substantial (here, 20 components) and can be arbitrary big. Next, the PCs are used as input variables in the regression (\ref{eq:reg:P:PC}).  In order to estimate the model's parameters, LASSO method is applied. This approach enables calibration of the model even when the number of PCs is larger than the size of the averaging calibration window. Moreover, it shrinks the parameters toward zero and hence reduces the noise induced by redundant components. Finally, the predictions of all hours of day $d_f$ are calculated (\ref{eq:reg:P:forecast})  and transformed back into the original units (\ref{eq:reg:P:transform}).

The LASSO optimization algorithm depends on a parameter $\lambda$, which specifies the impact of the penalty function. Similar to LASSO averaging, we consider a log-scaled grid of 20 $\lambda$ and select the optimal value via IC.  The outcomes are denoted either by \textbf{LPCA($\lambda$)} or by \textbf{LPCA(BIC)}, \textbf{LPCA(AIC)}, \textbf{LPCA(HQC)}, respectively. 

Since the LPCA does not require any prior decision neither on the size of the calibration windows used for averaging (as in AW/WAW) nor is restrictive in terms of the number of PC components (as in PCA), it can be perceived as a fully automated method. Moreover, thanks to the orthogonality of the PCs, the estimation algorithm is faster than LASSO averaging.

\section{Result}
\label{sec:res}

We use the \emph{Mean Absolute Error} (MAE) for the full out-of-sample test period of $D=916$ days (i.e., 29.06.2017 to 31.12.2019, see Figure \ref{fig:data1} or \ref{fig:data2}) as the main evaluation criterion. 
\red{It is one of the most commonly used measures for evaluation of forecast accuracy. In the case of electricity markets, it reflects the average deviation of the revenue from selling 1 MWh from its expected level. }
In \red{this} paper, we consider two \red{MAE-based} measures:
\begin{align}
    \label{eqn:MAE}
MAE_{d}^{(i)} &= \frac{1}{24} \sum_{h=1}^{24} |\varepsilon_{d,h}^{(i)}| \\
MAE^{(i)} &= \frac{1}{D} \sum_{d=1}^{D} MAE_{d}^{(i)}
\end{align}
where $\varepsilon_{d,h}^{(i)} = P_{d,h} - \hat{P}_{d,h}^{(i)}$ is the forecast error for hour $h$ in day $d$, \red{obtained either with different lengths calibration window, $\tau$, or averaging methods}. The first measure, $MAE_{d}^{(i)}$ describes the forecast accuracy for a single day $d$ \red{and is used for statistical comparison between individual approaches}. Finally $MAE^{(i)}$ describes the overall performance \red{in} the whole out-of-sample period.

As an auxiliary measure, we define a percentage change \red{of} forecast accuracy \red{relative} to the results of a model with the longest considered calibration window, it is 728-day \red{($MAE^{(\text{728})}$)}.
\begin{equation}
\label{eqn:chng}
\text{\% chng}_{i} = \frac{MAE^{(i)} - MAE^{(\text{728})} }{MAE^{(\text{728})}} \times 100\%,
\end{equation}
The relative change in the accuracy of a given model shows how different the model is from the usual approach of taking as long calibration windows as possible. Note that the positive sign of the measure indicates that a given model is worse than the benchmark, while the negative value appears when a given model outperforms the longest window approach. 

Given a number of datasets, it is hard to rank the models' accuracy. To solve this issue, we use a mean of the $\text{\% chng}_{i}$ over four datasets to obtain the final ranking.

\begin{equation}
\label{eqn:mpdb}
\text{m.p.d.b.}_{i} = \frac{1}{4} \sum_{m=1}^{4}  \text{\% chng}^{\red{m}}_{i},
\end{equation}
where \red{$m$} indicates one of four datasets (EPEX, NP, OMIE, PJM).

The obtained MAE values can be used to provide a ranking of \red{forecasts}. Unfortunately, they do not allow to draw statistically significant conclusions on the outperformance of \red{one prediction over the another}. Therefore, the \emph{conditional predictive ability} (CPA) test of \cite{gia:whi:06} is used to compare competitive outcomes. The test statistic is computed using the vector of average daily $MAE_d$:

\begin{equation}
\Delta_{i,j,d} = MAE_{d}^{(i)} - MAE_{d}^{(j)},
\end{equation}
\noindent
For each pair \red{$(i,j)$},  the $p$-value of the CPA test \red{is computed}.

\subsection{Individual forecasts}
\label{ssec:point}

\begin{figure}[tbp]
	\centering
	\includegraphics[width=.95\textwidth]{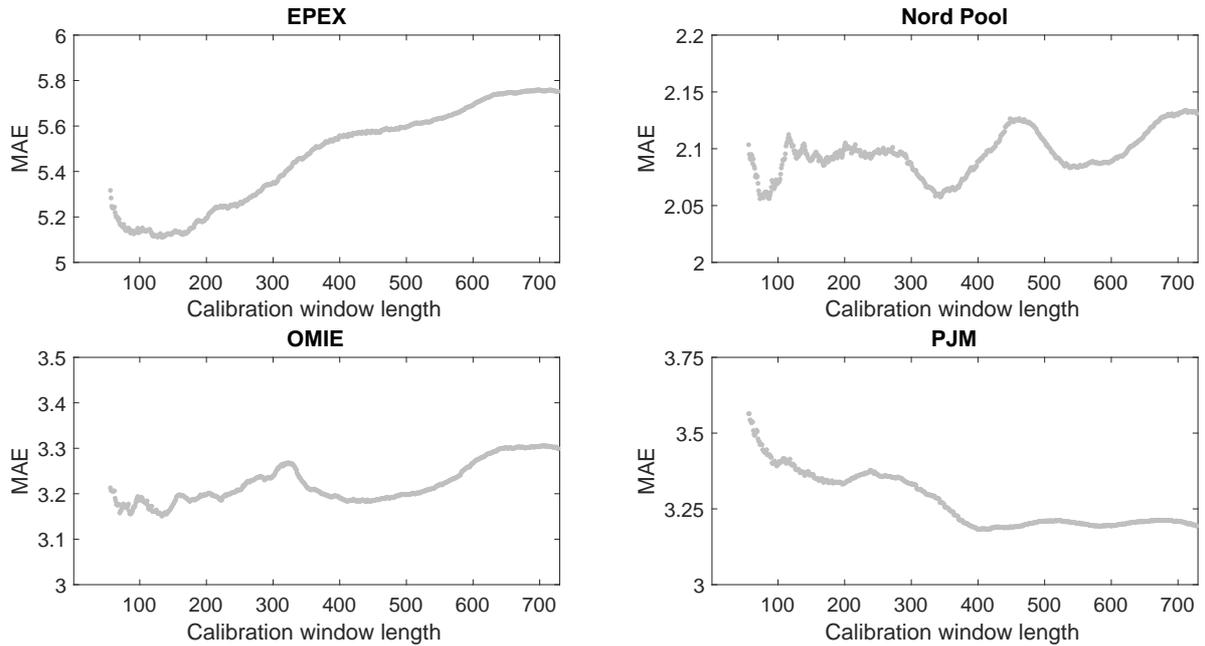}
	\caption{The \emph{Mean Absolute Errors} (MAE) for the EPEX, Nord Pool, OMIE and PJM datasets from the period 29.06.2017 to 31.12.2019 as a function of the calibration window length ranging from 56 to 728 days.}
	\label{fig:MAE}
\end{figure}

The performance of individual forecasts are presented in  Figure \ref{fig:MAE}, which shows the values of MAE for different calibration window lengths in the four analyzed markets. It could be observed that the strategy for selecting the optimal size of the calibration window differs between the datasets.
For some markets, such as EPEX, the longer the calibration window we take, the worse the autoregressive model performs. For others, such as PJM, it is beneficial to use long samples to estimate the model parameters. Finally, for Nord Pool and OMIE, the MAE plots are not monotonic and it is difficult to make the optimal decision. Hence, the results confirm the previous findings of \cite{hub:mar:wer:19} and \cite{mar:ser:wer:18} and prove that it is impossible to  ex-ante choose optimally the length of the  calibration window size.

Table \ref{tab:point} presents the detailed results for three selected window sizes: 56 days (8 weeks), 364 days (a year) and 728 days (2 years). They are next compared with the benchmark, which is the longest available calibration window. The outcomes are augmented with the results for the optimal window size, which is selected ex post and hence is not available for real-time usage. The results indicate that the selection of the calibration window length may have a great impact on the forecast accuracy. The gains from its proper choice reach up to $12.527\%$ (EPEX market).

\begin{table}[t]
	\centering
 	\caption{Mean absolute errors (MAE) and the percentage change (\%chng) compared to 'Simple average' benchmark of the price forecast for whole 916-days out-of-sample period from 29.06.2017 to 31.12.2019. Results are presented for selected calibration window length of 56, 364 or 728 days.}
	\label{tab:point}
	\scriptsize
	\begin{tabular}{cc c cc c cc c cc c cc}
		\toprule
		& && \multicolumn{2}{c}{\textbf{EPEX}}  && \multicolumn{2}{c}{\textbf{NP}} && \multicolumn{2}{c}{\textbf{OMIE}} && \multicolumn{2}{c}{\textbf{PJM}} \\[2pt]
		\multicolumn{2}{c}{\textbf{Calib. window length}} && \textbf{MAE} &  \textbf{\%chng}  && \textbf{MAE} &  \textbf{\%chng}  && \textbf{MAE} &  \textbf{\%chng} && \textbf{MAE} &  \textbf{\%chng} \\
		\midrule
\multicolumn{2}{c}{56}	&&	5.339	& -9.126 \% &&	2.210	& -1.365 \% &&	3.181	& -2.545 \% &&	3.674	& 11.075 \% \\[3pt]
\multicolumn{2}{c}{364}	&&	5.599	& -4.695 \% &&	2.163	& -3.45 \% &&	3.141	& -3.767 \% &&	3.352	& 1.317 \% \\[3pt]
\multicolumn{2}{c}{728}	&&	5.875	& 0 \% &&	2.241	& 0 \% &&	3.264	& 0 \% &&	3.308	& 0 \% \\[3pt]
\multicolumn{2}{c}{best}	&&	5.139	& -12.527 \% &&	2.159	& -3.651 \% &&	3.100	& -5.04 \% &&	3.299	& -0.28 \% \\[3pt]
\bottomrule
		\end{tabular}
		
\end{table}

\subsection{Averaging results}

Tables \ref{tab:results} and \ref{tab:ful_aut_results} present MAE and \%chng results for the forecasts obtained with different averaging techniques. 
\red{Here, two approaches are evaluated separately: semi-automated and fully-automated. In the first group of methods, arbitrary decisions of researchers about the number of components to be averaged are allowed. Moreover, the penalty parameter $\lambda$ in LASSO method is pre-defined for the whole sample. In the second group, the methods are fully automated, which means that the forecaster is not involved in the averaging process. }

\subsubsection{Semi-automated averaging methods}

Let us first analyze the outcomes of semi-automated approaches, in which the researcher decides a prior on the selection of forecasts used for averaging. In all considered methods, the inputs are chosen once for the whole evaluation period and do not adjust as the calibration and averaging windows move. The results are reported in Table \ref{tab:results}. First, the outcomes of AW and WAW methods are presented that are based only on a small subset of individual point forecasts (three short and three long windows). It can be observed that both approaches yield results, which are far better than the benchmark. By averaging forecasts stemming from just six different calibration windows, the MAE is reduced by more than $10\%$ for EPEX, NP and OMIE and at least $3\%$ for PJM. When both methods are compared, it can be observed that the weighted approach is better than AW, which assigns equal weights for all predictions.

Next, the error measures for LASSO, PCA, and LPCA with parameters selected \textit{ad hoc}, based on existing literature and experience, are presented. For each method, first three rows show outcomes for exemplary specifications described either by the number of components, $k$, in PCA(k) or $\lambda$ in LASSO($\lambda$) and LPCA($\lambda$).  The forth row reports results for the best \textit{ex-post} value of these parameters. The outcomes confirm that using forecast averaging techniques is beneficial. Similar to AW/WAW, all three methods enables substantial reduction of MAE, with the following specifications being the best: LASSO(10$^0$), PCA(5) and LPCA(10$^{-2}$).

When LASSO averaging scheme is considered, it can be observed that the results depend strongly on the parameter $\lambda$. There are substantial differences between LASSO(10$^{-2}$) and LASSO(10$^0$), which reach 18.922\% of the benchmark MAE for NP and 12.906\% for PJM. Moreover, LASSO(10$^{-2}$) is the worst of the averaging schemes and provides predictions less accurate than the 2-year calibration window for NP and PJM markets.

Performance of PCA is more robust to selection of the specification parameter, $k$. The relation between MAE of PCA and the number of components, $k$, is non-monotonic. First, as the number of PCs increases, the forecasts become more accurate. As it reaches the optimal level of $k$,  additional components introduce noise and lead to a rise of MAE. Hence, increasing the number of components does not improve the overall performance of the method.

When the results of LPCA are analyzed, it could be observed that LPCA inherits the positive features of both PCA and LASSO and reduces their weaknesses. Similar to PCA, LPCA it is robust to the choice of the tuning parameter, $\lambda$. On the other hand, it allows to use a large number of components without a loss of efficiency because  LASSO allows to reduce the parameter space.

Finally, when the LASSO(best), PCA(best) and LPCA(best) are compared, LPCA and LASSO are both the best in two out of four markets with the PCA scheme never reaching the top of the podium. The aggregated results, summarized by m.p.d.b., confirm that LPCA yields the most accurate predictions among any alternatives.

The results for non-automated averaging approaches can be summarised by following conclusions: 
\begin{itemize}
    \item Almost all averaging approaches (except LASSO($10^{-2}$) outperform the 'longest window' model by a large margin, often even higher than 10\%
    \item The most accurate forecast can be obtained with LASSO and LPCA, both are the best for two out of four datasets.
    \item The performance of LASSO depends strongly on $\lambda$, wheres PCA and LPCA are more robust to the choice of the specification parameters.
    \item The idea of AW and WAW, introduced by \cite{hub:mar:wer:19} and \cite{mar:ser:wer:18}, performs very well, however it can be outperformed by more sophisticated approaches
\end{itemize}

\begin{table}[H]
	\centering
 	\caption{Mean absolute errors (MAE) and the percentage change (\%chng) compared to 'Simple average' benchmark of the price forecast for whole 916-days out-of-sample period from 29.06.2017 to 31.12.2019. In this panel we report the result obtained with averaging setups which are depended on the forecaster knowledge/intuition. Note that in each column the best result is bolded.}
	\label{tab:results}
	\scriptsize
	\begin{tabular}{cc c cc c cc c cc c cc c c}
		\toprule
		& && \multicolumn{2}{c}{\textbf{EPEX}}  && \multicolumn{2}{c}{\textbf{NP}} && \multicolumn{2}{c}{\textbf{OMIE}} && \multicolumn{2}{c}{\textbf{PJM}} &&\\[2pt]
		\multicolumn{2}{c}{\textbf{Averaging}} && \textbf{MAE} &  \textbf{\%chng}  && \textbf{MAE} &  \textbf{\%chng}  && \textbf{MAE} &  \textbf{\%chng} && \textbf{MAE} &  \textbf{\%chng}&& \textbf{m.p.d.b.} \\
		\midrule
\multicolumn{2}{c}{AW}	&&	5.059	& -13.895 \% &&	1.970	& -12.101 \% &&	2.917	& -10.629 \% &&	3.206	& -3.099 \% &&-9.931 \%\\[3pt]
\multicolumn{2}{c}{WAW}	&&	5.014	& -14.65 \% &&	1.966	& -12.264 \% &&	2.913	& -10.755 \% &&	3.204	& -3.148 \% &&-10.204 \%\\[3pt]
\hline\\
\multicolumn{2}{c}{LASSO(10$^{-2}$)}	&&	5.416	& -7.822 \% &&	2.408	& 7.464 \% &&	3.029	& -7.216 \% &&	3.657	& 10.534 \% && 0.740 \%\\[3pt]
\multicolumn{2}{c}{LASSO(10$^{-1}$)}	&&	4.954	& -15.671 \% &&	2.018	& -9.954 \% &&	2.886	& -11.575 \% &&	3.255	& -1.595 \% && -9,699 \%\\[3pt]
\multicolumn{2}{c}{LASSO(10$^{0}$)}	&&	\textbf{4.962}	& \textbf{-15.536} \% &&	1.984	& -11.458 \% &&	\textbf{2.893}	& \textbf{-11.356} \% &&	3.230	& -2.372 \%&& -10.180 \% \\[3pt]
\multicolumn{2}{c}{LASSO(best)}	&&	4.924	& -16.182 \% &&	1.963	& -12.37 \% &&	2.872	& -12.023 \% &&	3.200	& -3.268 \% && -10.961 \%\\[3pt]
\hline\\
\multicolumn{2}{c}{PCA(1)}	&&	5.030	& -14.38 \% &&	2.025	& -9.612 \% &&	2.963	& -9.21 \% &&	3.269	& -1.195 \% && -8.599 \%\\[3pt]
\multicolumn{2}{c}{PCA(5)}	&&	5.007	& -14.771 \% &&	1.980	& -11.647 \% &&	2.913	& -10.766 \% &&	3.220	& -2.672 \% &&-9.964 \%\\[3pt]
\multicolumn{2}{c}{PCA(20)}	&&	5.080	& -13.524 \% &&	2.069	& -7.663 \% &&	2.944	& -9.803 \% &&	3.278	& -0.915 \% && -7.976 \% \\[3pt]
\multicolumn{2}{c}{PCA(best)}	&&	4.965	& -15.495 \% &&	1.969	& -12.103 \% &&	2.913	& -10.766 \% &&	3.210	& -2.972 \% && -10.334 \% \\[3pt]
\hline\\
\multicolumn{2}{c}{LPCA(10$^{-3}$)}	&&	4.998	& -14.93 \% &&	2.022	& -9.745 \% &&	2.914	& -10.728 \% &&	3.244	& -1.942 \% && -9.336 \%\\[3pt]
\multicolumn{2}{c}{LPCA(10$^{-2}$)}	&&	4.979	& -15.253 \% &&	\textbf{1.970}	& \textbf{-12.064} \% &&	2.904	& -11.045 \% &&	\textbf{3.202}	& \textbf{-3.209} \% &&\textbf{-10.393} \%\\[3pt]
\multicolumn{2}{c}{LPCA(10$^{-1}$)}	&&	5.107	& -13.066 \% &&	2.058	& -8.151 \% &&	3.014	& -7.663 \% &&	3.270	& -1.149 \% && -7.507 \%\\[3pt]
\multicolumn{2}{c}{LPCA(best)}	&&	4.970	& -15.406 \% &&	1.961	& -12.473 \% &&	2.893	& -11.361 \% &&	3.197	& -3.369 \% && -10.652 \% \\[3pt]
		\bottomrule
	\end{tabular}
\end{table}

\subsubsection{Fully automated averaging methods}

In this  article, four fully automated forecast averaging methods are considered. These are approaches, which do not require any expert knowledge to select the inputs used for forecast averaging or to specify parameters such as the number of components, $k$, in PCA and a value of LASSO tuning parameter, $\lambda$. The results are presented in Table \ref{tab:ful_aut_results}, which similar to Table \ref{tab:results} shows MAE forecast accuracy measure and \%chng.

The first method is a simple average. It is an automated approach because it does not require any pre-selection of predictions used for pooling. This method provides forecasts, which are far better than the benchmark. It reduces MAE by 1.149\%-- 10.221\%, which is slightly less than in AW/WAW case.

Next, three methods: LASSO, PCA and LPCA are analyzed. Unlike the previous section, here the tuning parameters: $k$ and $\lambda$, are selected with Information Criteria (AIC, BIC and HQ). This modification has two major advantages. First, it does not require a prior knowledge on the specification of these methods in a particular application. Hence, it can be easily used for predicting prices of other commodities or for any other forecasting exercises. Second, the parameters can evolve as new data arrives and adjust to the market situation.

First, it could be noticed that LASSO method is sensitive to the choice of IC. For AIC, it provides forecasts, which are less accurate than a benchmark for three out of four analyzed markets. For PJM, the loss of accuracy exceeds 20\%. Even for EPEX market, for which the gains are the highest, LASSO(AIC) is only slightly better than the predictions obtained with the longest calibration window. Moreover, LASSO(HQC), although better than LASSO(AIC), does not provide satisfactory results. It improves the predictions for EPEX, NP and OMIE but worsens them for PJM by more than 7\%. Only LASSO(BIC) gives results, which are consistently better than the benchmark.

Similar to LASSO, the performance of LPCA approach depends on the choice of IC. In this case, the differences between ICs are less pronounced, with LPCA(BIC) providing the most accurate predictions. Hence, as well as for the standard LASSO, also for LPCA it is BIC that should be used for selecting the parameter $\lambda$. It is worth noting that all three LPCA methods produce the best forecasts in terms of MAE for EPEX, NP and PJM markets. They are outperformed only by LASSO(BIC) in OMIE case.

In the case of the PCA method, it is hard to choose the clear winner between different ICs. For each dataset, a different approach provides the most accurate results. The differences, however, are not substantial, so the optimal number of PCs can be successfully selected via any of the considered ICs.  Although the most robust, the approach is never the best choice in terms of MAE accuracy, as it is outperformed by either LPCA or LASSO.

The last column of Table \ref{tab:ful_aut_results} presents m.p.d.b, the aggregated measure of forecast accuracy. The outcomes show that well-designed averaging models can outperform the most popular approach of an arithmetic mean. Moreover, they confirm previous findings obtained using semi-automated methods and indicate that LPCA reduces MAE more than other averaging approaches.

\begin{table}[b!]
	\centering
 	\caption{Mean absolute errors (MAE) and the percentage change (\%chng) compared to 'Simple average' benchmark of the price forecast for whole 916-days out-of-sample period from 29.06.2017 to 31.12.2019. Presented results correspond to the fully automated approaches of averaging technique. Note that in each column the best result is bolded.}
	\label{tab:ful_aut_results}
	\scriptsize
	\begin{tabular}{cc c cc c cc c cc c cc c c}
		\toprule
		& && \multicolumn{2}{c}{\textbf{EPEX}}  && \multicolumn{2}{c}{\textbf{NP}} && \multicolumn{2}{c}{\textbf{OMIE}} && \multicolumn{2}{c}{\textbf{PJM}} &&\\[2pt]
		\multicolumn{2}{c}{\textbf{Fully automated}} && \textbf{MAE} &  \textbf{\%chng}  && \textbf{MAE} &  \textbf{\%chng}  && \textbf{MAE} &  \textbf{\%chng} && \textbf{MAE} &  \textbf{\%chng} && \textbf{m.p.d.b.}\\
		\midrule
\multicolumn{2}{c}{simple average}	&&	5.275	& -10.221 \% &&	2.068	& -7.706 \% &&	3.021	& -7.448 \% &&	3.270	& -1.149 \% &&-6.631 \%\\[3pt]
\hline\\
\multicolumn{2}{c}{LASSO(AIC)}	&&	5.853	& -0.379 \% &&	2.304	& 2.81 \% &&	3.285	& 0.632 \% &&	3.998	& 20.866 \% && 5.982 \%\\[3pt]
\multicolumn{2}{c}{LASSO(BIC)}	&&	5.005	& -14.811 \% &&	1.989	& -11.235 \% &&	\textbf{2.898}	& \textbf{-11.206} \% &&	3.259	& -1.48 \% && -9.683 \%\\[3pt]
\multicolumn{2}{c}{LASSO(HQC)}	&&	5.221	& -11.131 \% &&	2.084	& -6.971 \% &&	2.968	& -9.071 \% &&	3.542	& 7.069 \% && -5.026 \%\\[3pt]
\hline\\
\multicolumn{2}{c}{PCA(AIC)}	&&	5.012	& -14.694 \% &&	2.014	& -10.105 \% &&	2.949	& -9.664 \% &&	3.249	& -1.791 \% &&-9.064 \%\\[3pt]
\multicolumn{2}{c}{PCA(BIC)}	&&	5.018	& -14.59 \% &&	1.987	& -11.342 \% &&	2.953	& -9.54 \% &&	3.252	& -1.691 \% &&-9.291 \%\\[3pt]
\multicolumn{2}{c}{PCA(HQC)}	&&	5.005	& -14.806 \% &&	2.004	& -10.578 \% &&	2.945	& -9.792 \% &&	3.251	& -1.737 \% &&-9.228 \%\\[3pt]
\hline\\
\multicolumn{2}{c}{LPCA(AIC)}	&&	4.947	& -15.796 \% &&	1.988	& -11.26 \% &&	2.931	& -10.2 \% &&	3.221	& -2.626 \% &&-9.971 \%\\[3pt]
\multicolumn{2}{c}{LPCA(BIC)}	&&	\textbf{4.923}	& \textbf{-16.212} \% &&	\textbf{1.979}	& \textbf{-11.689} \% &&	2.924	& -10.426 \% &&	\textbf{3.217}	& \textbf{-2.759} \%&&  \textbf{-10.271} \% \\[3pt]
\multicolumn{2}{c}{LPCA(HQC)}	&&	4.927	& -16.137 \% &&	1.988	& -11.292 \% &&	2.932	& -10.182 \% &&	3.221	& -2.624 \% &&-10.059 \%\\[3pt]
		\bottomrule
	\end{tabular}
\end{table}

\begin{figure}[tbp]
	\centering 
	\includegraphics[height = 5cm]{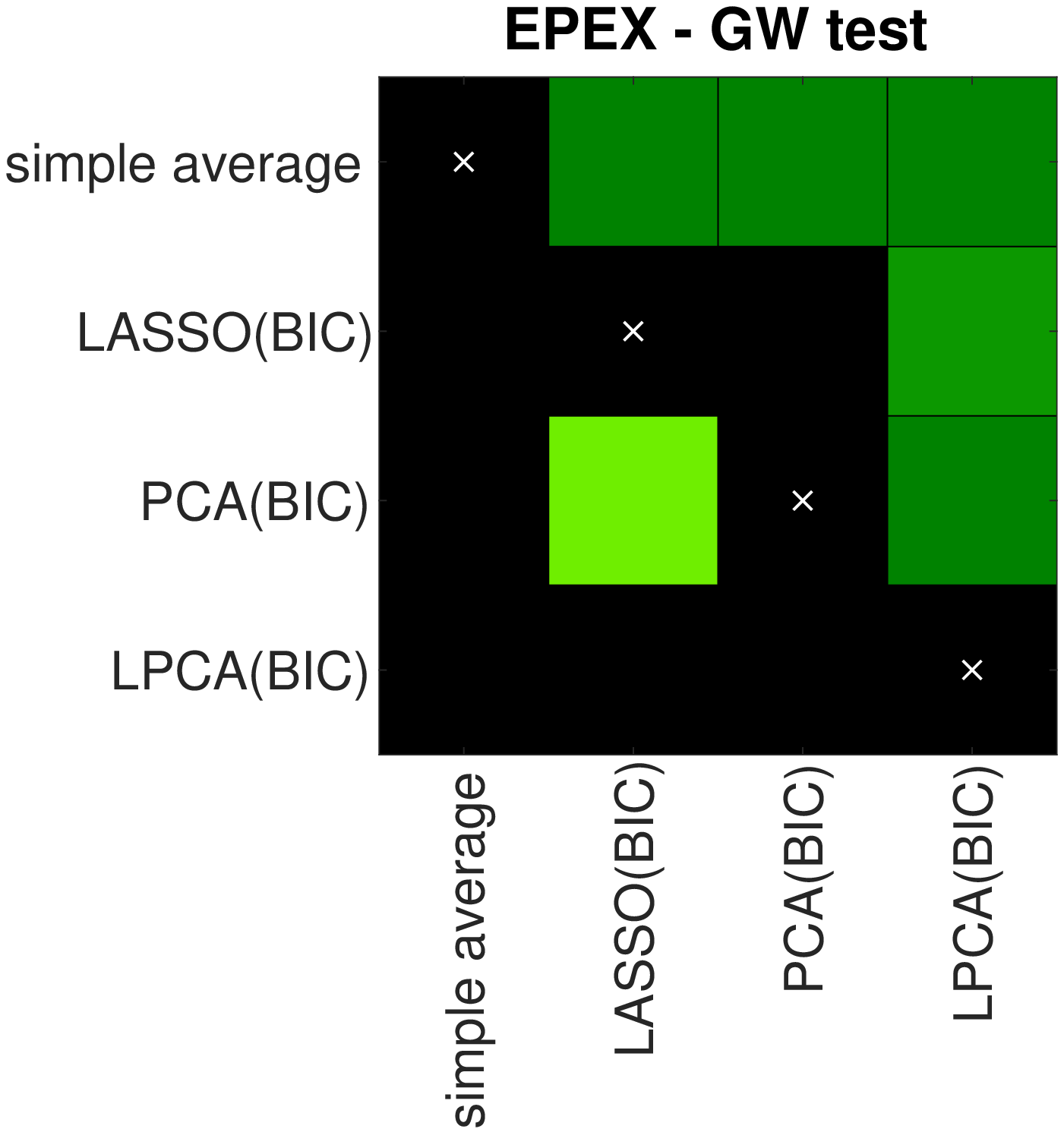}
	\hspace*{.2cm}
	\includegraphics[height = 5cm]{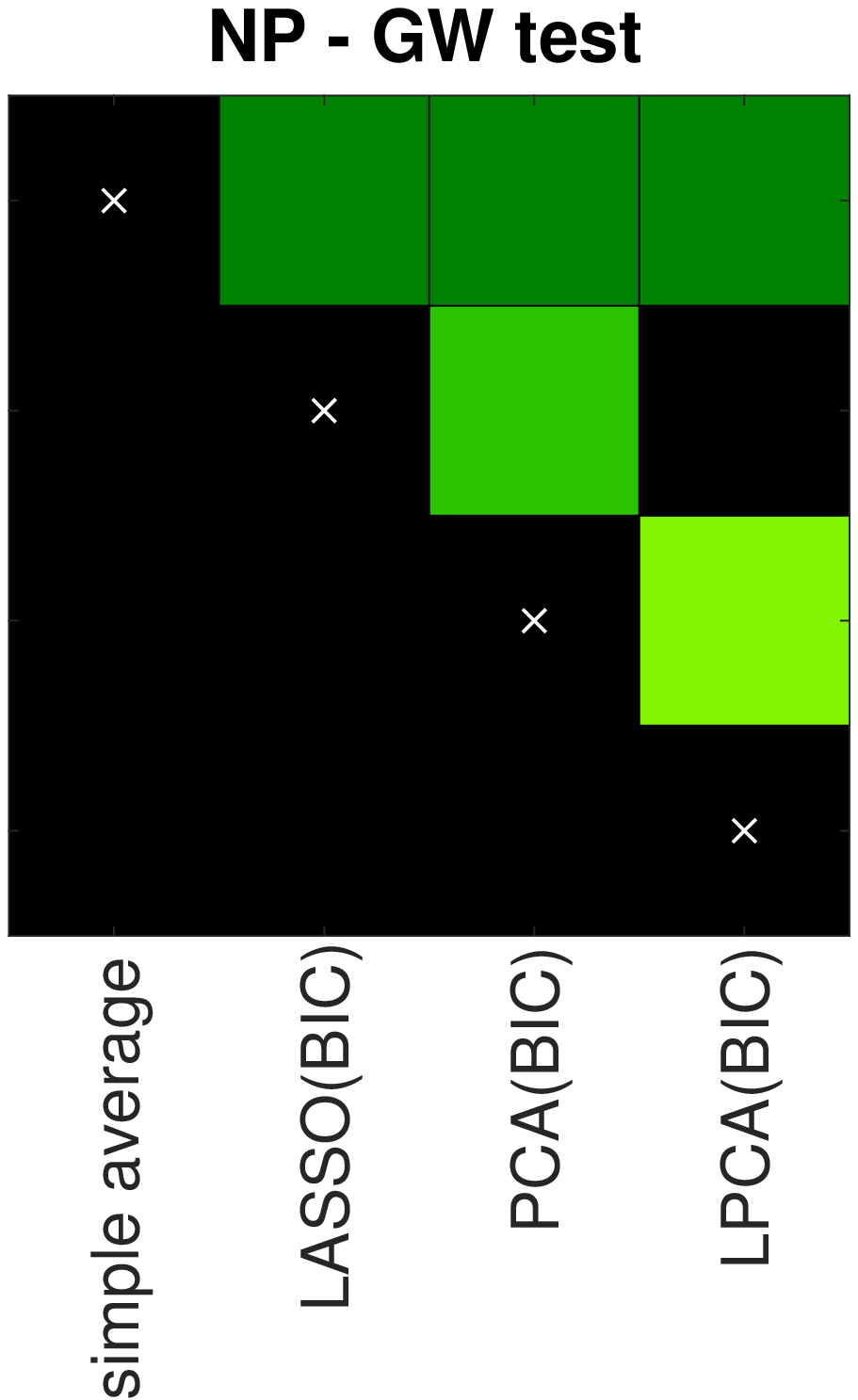}
	\hspace*{.2cm}
	\includegraphics[height = 5cm]{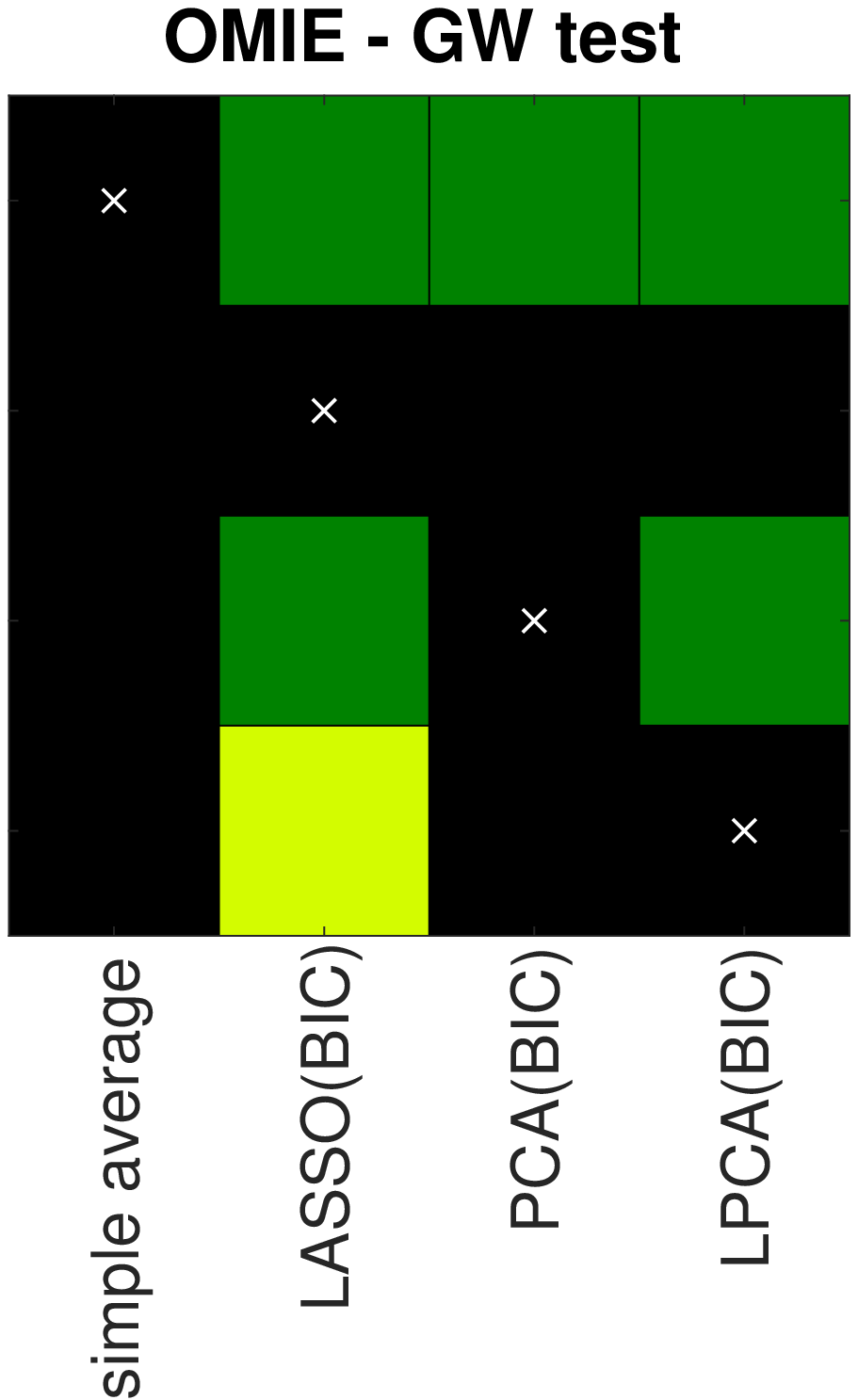}
	\hspace*{.2cm}
	\includegraphics[height = 5cm]{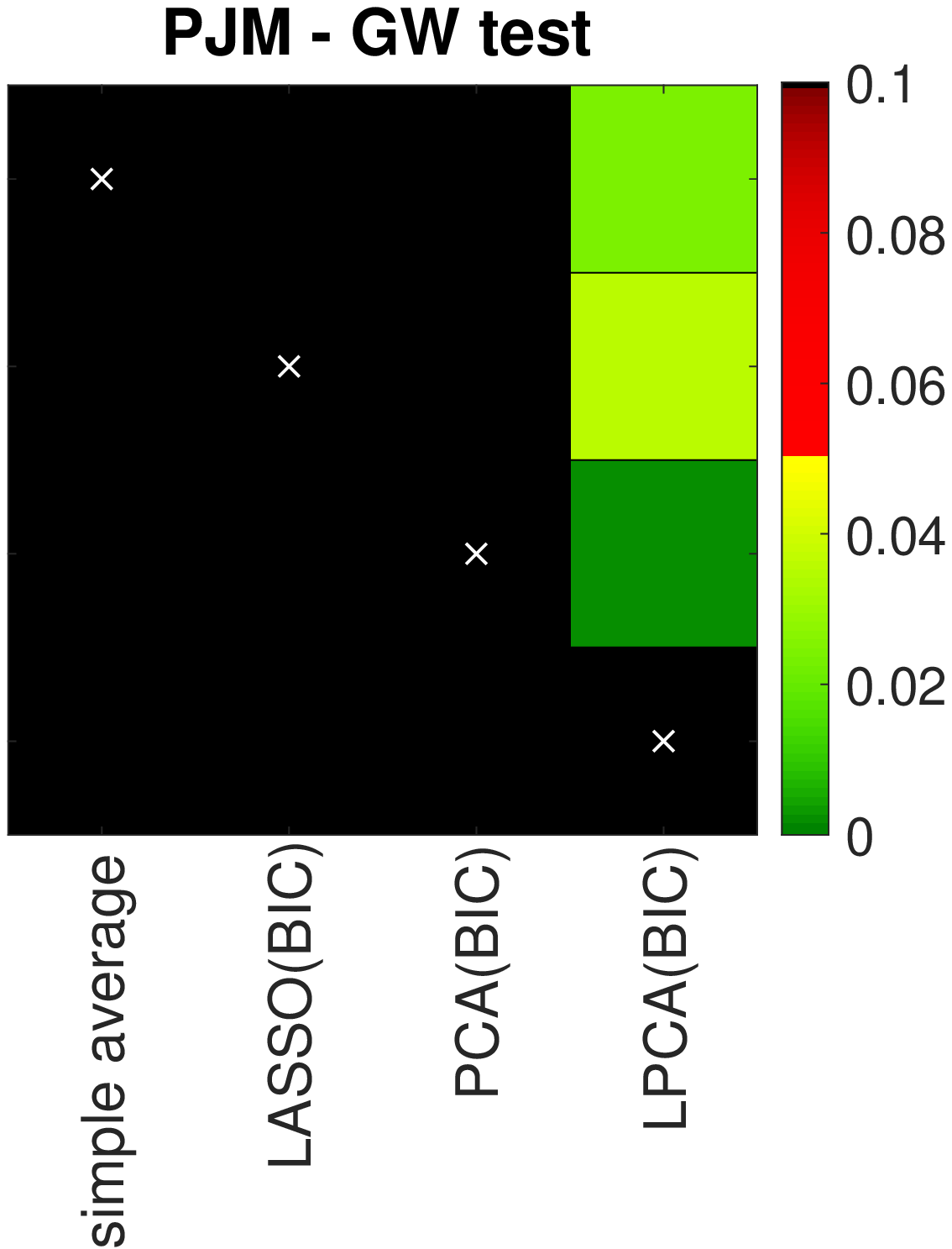}
	\caption{Results of the conditional predictive ability (CPA) test of \cite{gia:whi:06} for forecasts of selected models for the EPEX (\emph{left}), Nord Pool (\emph{left center}), OMIE (\emph{right center}) and PJM (\emph{right}) datasets. We use a heat map to indicate the range of the $p$-values -- the closer they are to zero ($\rightarrow$ dark green) the more significant is the difference between the forecasts of a model on the $X$-axis (better) and the forecasts of a model on the $Y$-axis (worse).}
	\label{fig:GW}
\end{figure}

To formally investigate the advantages of using our newly proposed averaging method, we apply the Conditional Predictive Ability (CPA; see \cite{gia:whi:06}) test for significant differences in the forecasting performance. The outcomes are presented in Figure \ref{fig:GW}, on which a non-black square indicates that the forecasts of the model on the $X$-axis are statistically more accurate than the forecasts of a model on the $Y$-axis. The results confirm the previous findings and show the LPCA extension of the standard PCA approach significantly outperforms other methods, in particular simple mean and PCA, for each considered dataset. What is more, it is two out of four times significantly better compared to the LASSO and never worse.

Finally, it could be noticed that the simple average is almost every time outperformed by other averaging approaches. This result shows that the arithmetic mean is useful as a benchmark for the newly introduced methodology, however, it should not be treated as a golden standard.

To sum up:
\begin{itemize}
    \item Almost all averaging approaches (except LASSO(AIC) and LASSO(HQC)) can easily beat the 'longest window' model by a large margin
    \item Among the forecasts based on LASSO or LPCA methods, the most accurate results are obtained with BIC.
    \item The PCA method is the most robust to the choice of IC. None of the ICs dominates and all of them provide similar results.
    \item Overall, the best result can be obtained with LPCA(BIC). 
\end{itemize}

\subsection{Discussion}

\begin{table}[tb]
	\centering
 	\caption{Comparison of averaging methods based on m.p.d.b across different specifications: choice of tuning parameters, $k$ or $\lambda$, for semi automated  or IC for fully automated approaches; MAD - Mean Absolute Deviation}
	\label{tab:aggregation}
	\scriptsize
	\begin{tabular}{c c c c c c c}
		\toprule
		& & \textbf{Top}& &  \textbf{Mean}& &  \textbf{MAD}\\
		\midrule
		& \multicolumn{6}{c}{\textbf{Semi automated methods}} \\
		\midrule
LASSO& & -10.961 \% & &  -5.981 \% & &  5.165 \%\\[3pt]
PCA& & -10.334 \% & &  -9.232 \% & &  0.614 \%\\[3pt]
LPCA & & -10.652 \% & &  -9.582 \% & &  0.802 \%\\[3pt]
	\midrule
		& \multicolumn{6}{c}{\textbf{Fully automated methods}} \\
	\midrule
	LASSO& & -9.683\% & &  -2.909\% & & 5.927 \%\\[3pt]
PCA& & -9.291\% & &  -9.194\% & &  0.087\%\\[3pt]
LPCA & & -10.271\% & &  -10.100\% & &  0.114\%\\[3pt]
	\bottomrule 
		\end{tabular}
\end{table}

\red{In this research, a performance of different averaging schemes based on forecasts obtained with different calibration windows is analyzed. It is shown}
that it is beneficial to pool predictions even when they come from a single model. \red{A} large number of individual forecasts available for averaging becomes both the advantage and the main issue of this idea, which makes it difficult to fully automate the computations. Here, two approaches are explored that are based on information and parameter space reduction. PCA method allows to summarize the \red{data described by} a panel of forecasts with a relatively small set of orthogonal components, whereas LASSO shrinks the model's parameters toward zero and hence increase the estimation efficiency. This research demonstrates that the application of both approaches can result in a substantial increase of forecast accuracy. Unfortunately, the methods are burdened with the uncertainty associated with the choice of tuning parameters. The dependence of the results on this selection is illustrated in Table \ref{tab:aggregation}, which shows the best outcomes in terms of m.p.d.b together with the mean and the Mean Absolute Deviation (MAD) of m.p.d.b across different specifications. The results indicate that  although LASSO($10^0$) and LASSO(BIC) are among the best forecast averaging approaches, the LASSO method is sensitive to the selection of the tuning parameter and the IC. Its average m.p.d.b. is slightly less than 6\% and 3\% for semi- and fully automated approaches, respectively. At the same time, LPCA improves forecasts by 9.582\% and 10.1\%\red{, respectively}. Moreover, PCA and LPCA methods are characterized by low values of MAD, which are far smaller than in LASSO case.

The difference in performance of LASSO, PCA, and LPAC forecast averaging methods results from their construction. When  PCA approach is considered, it should be underlined that components used for averaging are orthogonal to each other and hence enables efficient estimation of (\ref{eq:reg:P:forecast}) parameters. However, unlike  LPCA, this approach includes all PCs from 1 to $k$ in the regression. Application of LASSO to (\ref{eq:reg:P:forecast}) allows to reduce the parameter space. The method does not only eliminate the insignificant components, but also shrinks the weights corresponding to less important variables. 
\cite{uni:wer:18} compared LASSO with two-step procedure  including variable selection via LASSO and estimating weights (of selected variables) via OLS. It turned out that LASSO significantly outperforms two-step procedure. Similar situation can be observed also in our research. The limited study presented in Table \ref{tab:PCA_BIC} shows that applying the two-step procedure does not improve (on average) the forecast accuracy compared to PCA(BIC). This indicates that the shrinkage is even more important than selection in our task. The regularization improves the averaging accuracy not because it allows to better select the number of PCs, but the LASSO shrunken weights are better to use in this setup.

\begin{table}[tb]
	\centering
 	\caption{Mean absolute errors (MAE) and the percentage change (\%chng) compared to 'Simple average' benchmark of the price forecast for whole 916-days out-of-sample period from 29.06.2017 to 31.12.2019. Results are presented to compare LASSO and two-step procedure proposed by \cite{uni:wer:18}}
	\label{tab:PCA_BIC}
	\scriptsize
	\begin{tabular}{cc c cc c cc c cc c cc c c}
		\toprule
		& && \multicolumn{2}{c}{\textbf{EPEX}}  && \multicolumn{2}{c}{\textbf{NP}} && \multicolumn{2}{c}{\textbf{OMIE}} && \multicolumn{2}{c}{\textbf{PJM}} \\[2pt]
		\multicolumn{2}{c}{\textbf{Averaging}} && \textbf{MAE} &  \textbf{\%chng}  && \textbf{MAE} &  \textbf{\%chng}  && \textbf{MAE} &  \textbf{\%chng} && \textbf{MAE} &  \textbf{\%chng} && m.p.d.b\\
		\midrule
\multicolumn{2}{c}{PCA(BIC)}	&&	5.018	& -14.590 \% &&	1.987	& -11.342 \% &&	2.953	& -9.54 \% &&	3.252	& -1.691 \% && -9.291\\[3pt]
\multicolumn{2}{c}{LPCA(BIC)}	&&	4.923	& -16.212 \% &&	1.979	& -11.689 \% &&	2.924	& -10.426 \% &&	3.217	& -2.759 \% && -10.272\\[3pt]
\multicolumn{2}{c}{2-step(BIC)}	&&	5.057	& -13.919 \% &&	2.019	& -9.901 \% &&	2.959	& -9.348 \% &&	3.238	& -2.125 \% && -8.823\\[3pt]
	\bottomrule
		\end{tabular}
\end{table}

Finally, when LASSO and  LPCA methods are compared, it could be noticed that LASSO has many more inputs than  LPCA. Extracting information from the panel of forecasts via the PC reduces the dimension of the regression. Moreover, unlike the PCs, the individual forecasts are highly correlated and almost co-linear. Due to these features, LASSO is more sensitive to specification of the tuning parameter. Moreover, the CPU needed to compute the forecast with LASSO is 900-times higher compared to the time needed to perform LPCA.

\section{Conclusions}
\label{sec:conclusion}

In this paper, a novel approach for point forecast pooling is presented, which combines both LASSO estimation method and PCA averaging scheme introduced by \cite{mac:uni:ser:20}. PCA allows to summarize the information included in a panel of forecasts with a relatively small set of orthogonal components, whereas LASSO shrinks the model's parameters toward zero and hence increase the estimation efficiency. The performance of the approach is evaluated on datasets from four major energy markets. Following \cite{mar:ser:wer:18} and \cite{hub:mar:wer:19}, the point predictions used for pooling stem from a single ARX-type model calibrated to windows of different sizes. The forecasts are evaluated with MAE and the results are presented relative to the outcomes obtained with the longest available calibration window, which includes two years of observations.

The results confirm previous findings of \cite{mar:ser:wer:18} and \cite{mac:uni:ser:20} that the longest estimation window does not necessarily lead to the most accurate predictions. Hence, it is not possible to select a prior optimal length of the sample used for calibration. At the same time, averaging algorithms can substantially reduce  MAE and improve the forecast accuracy relative to the benchmark, by -6.631\% for a simple average and  -10.271\% for LPCA(BIC) approach.

When the forecast averaging methods are considered, the outcomes indicate that fully automated approaches, which use Information Criteria to select an optimal specification, yield results which are significantly better than the benchmark or the simple average. The performance of the presented pooling methods depends, however, on applied IC. The outcomes show that BIC is the most robust choice, which leads to the lowest relative MAE for all approaches. The comparison of LASSO, PCA and LPCA allows to draw the following conclusions:

\begin{itemize}
    \item The PCA method is the most robust to the choice of IC, however, it reduces MAE less than the methods using LASSO
    \item LASSO is extremely sensitive to the choice of the tuning parameter and IC
    \item Overall LPCA outperforms other approaches: it improves the forecast accuracy the most and is relatively robust to the selection of the tuning parameter
\end{itemize}

The LPCA approach, which combines LASSO with PCA, is proved to be successfully in forecasting day-ahead electricity prices.
\red{This research could be viewed as a first step in mixing PCA with automated variable selection methods.
Future analysis may include more complex models such as elastic net, adaptive lasso, or neural network-based. Moreover, the research may be extended to}
interval and probabilistic forecasting and be applied to other commodity markets.

\section*{Acknowledgments}
This work was partially supported by the Ministry of Science and Higher Education (MNiSW, Poland) through Diamond Grant No. 0199/DIA/2019/48 (to BU) and National Science Centre (NCN, Poland) through SONATA BIS grant no. 2019/34/E/HS4/00060 (to KM)

\bibliographystyle{elsarticle-harv}
\bibliography{lasso}

\end{document}